\documentclass[pra,floatfix,10pt]{revtex4-1}
\usepackage{amsmath,amsfonts,amssymb,graphicx,color,bbm,psfrag,dsfont,slashed}
\usepackage[T1]{fontenc}
\usepackage[latin9]{inputenc}
\usepackage[colorlinks=true,citecolor=blue]{hyperref}

\newcommand{\beq}{\begin{equation}}
\newcommand{\eeq}{\end{equation}}
\newcommand{\bea}{\begin{eqnarray}}
\newcommand{\eea}{\end{eqnarray}}

\newcommand{\up}{\uparrow}
\newcommand{\down}{\downarrow}

\renewcommand{\b}[1]{\mathbf{ #1}}									%vettori grasetti
\newcommand{\cop}[1]{{#1^{\dagger}}\!}  						%creation opeator			
\newcommand{\aop}[1]{{#1}}

\begin{document}

\title{Effective control of chemical potentials 
by Rabi coupling with {\color{black} rf-fields} in ultracold mixtures}

\author{L. Lepori}
\email[correspondence at: ]{llepori81@gmail.com}
\affiliation{Dipartimento di Scienze Fisiche e Chimiche, Universit\`a dell'Aquila, via Vetoio,
I-67010 Coppito-L'Aquila, Italy}
\affiliation{INFN, Laboratori Nazionali del Gran Sasso, Via G. Acitelli,
22, I-67100 Assergi (AQ), Italy.}

\author{A. Maraga}
\affiliation{SISSA/ISAS,
via Bonomea 265, I-34136 Trieste, Italy.}

\author{A. Celi}
\affiliation{Institute for Theoretical Physics, University of Innsbruck, A-6020 Innsbruck, Austria and Institute for Quantum Optics and Quantum Information of the Austrian Academy of Sciences, A-6020 Innsbruck, Austria}
\affiliation{ICFO--Institut de Ciencies Fotoniques, The Barcelona Institute of Science and Technology, 08860 Castelldefels, Barcelona, Spain.}

\author{L. Dell'Anna}
\affiliation{Dipartimento di Fisica e Astronomia ``Galileo Galilei'' 
Universit\`a di Padova, I-35131 Padova, Italy}

\author{A. Trombettoni}
\affiliation{SISSA/ISAS and INFN, Sezione di Trieste,
via Bonomea 265, I-34136 Trieste, Italy}
\affiliation{CNR-IOM DEMOCRITOS Simulation Center, Via Bonomea 265, I-34136 Trieste, Italy.}

\begin{abstract}
We show that a linear term coupling the 
atoms of an ultracold binary mixture provides a simple 
method to induce an effective and tunable population imbalance between them. 
This term is easily realized by a Rabi coupling between 
different hyperfine levels of the same atomic species. 
The resulting effective imbalance holds for one-particle states dressed by the Rabi coupling and obtained diagonalizing the mixing matrix of the Rabi term. This way of controlling the chemical potentials applies for both bosonic and fermionic atoms and it allows also for spatially and temporally dependent imbalances.
As a first application, we show that, in the case of two attractive fermionic hyperfine levels with equal chemical potentials 
and coupled by the Rabi pulse, the same superfluid properties of an imbalanced binary mixture are recovered.  
We finally discuss the properties of $m$-species mixtures in the presence of SU($m$)-invariant interactions. 
\end{abstract}

\maketitle

\section{Introduction}

The average density of the 
particles is a central parameter in ultracold experiments~\cite{bloch08,lewenstein12}, and more in general for condensed matter phenomena.
Indeed, most of the physical systems  acquire different features 
when the interaction energy is varied by changing the number of particles $N$.
This fact arises already at the single-particle
level, determining for instance the shape of the Fermi surfaces, in turn related e.g. to 
conductance,
or to the critical properties of  Bose condensates \cite{revBE}. 
The phenomenon is even more pronounced once interactions are included as, for instance, 
it has been shown for superfluidity in {\color{black} fermionic} mixtures \cite{pethick}. In the presence
of a density imbalance between the pairing species, new types of superfluid phases, 
different from the standard BCS/BEC  ones, can appear \cite{pethick,marchetti,SR,chevy2011,shin2006,ketterle2006,hulet2006,salomon2010,bakr2016}. 
Other notable examples are the quantum Hall effect \cite{yoshioka}, 
where the filling deeply affects the nature of the ground state, or the transition superfluid-Mott insulator in the Bose-Hubbard model \cite{lewenstein12}.

In recent years, the experimental capability of controlling and measuring local particle densities has greatly improved, due to single site addressing  \cite{Bakr09,Weitenberg11}. However, typically in ultracold experiments one controls 
the density and not the chemical potential $\mu$. 
{\color{black} In order to go from the canonical 
description} to the grand-canonical one, one needs to exploit the relation between $N$ and $\mu$, generally a non-trivial task. Moreover, standard experiments with imbalanced mixtures are not always straightforward to realize, 
since often the phases separate or become unstable \cite{lewenstein12}. 
Therefore, {\color{black} designing} alternative methods to tune imbalances, 
even effectively, appears a very interesting task. This can turn out 
even more relevant if the chemical potential can be tuned in a 
space- or time-dependent way, which is not easy to realize controlling the 
atomic populations: for instance, this possibility may be crucial 
for the synthesis of unconventional superfluid pairings with nonzero momenta 
or for the creation of space-time defects.

In this paper, we provide a different method to induce and to control effective population imbalances in experiments involving atoms 
of a binary mixture linearly coupled between them, a major example being 
different hyperfine levels equally populated coupled by a {\color{black} Rabi rf-field.} 
The obtained imbalances hold for dressed one-particle states, 
obtained diagonalizing the Rabi term that couples the  hyperfine levels pairwise. The method works both for 
bosonic and fermionic atoms.
 The Rabi coupling in fermionic and bosonic 
mixtures is well studied for a variety of applications 
(see, e.g., \cite{Williams99,Gupta03,Smerzi03,Chen09,Pieri09,Barbiero16}) 
and widely used in 
ultracold atoms experiments \cite{lewenstein12,fallanibook,steck}. We therefore 
think that it would be straightforward to perform an experiment along the lines 
discussed in this paper {\color{black} for realizing} an effective imbalance between different hyperfine levels of ultracold mixtures.

In the presence of two-body interactions for the mixture, 
we argue that, under general conditions, the dynamics can be equally described in terms of the original states or of the dressed states, 
as the scattering processes do not destroy the coherence of the dressed states. A particularly interesting situation is when the two-body interactions do 
not depend on the involved {\color{black} hyperfine species} in the mixture. This is the case of the earth-alkaline like atoms, as $^{87}$Sr \cite{DeSalvo10,Tey10} or $^{173}$Yb \cite{Fukuhara07,Taie10}. These atomic species 
provide instances of SU($m$)-invariant mixtures 
(with $m=10$ and $6$, respectively). Quite recently, similar mixtures 
have been {\color{black} found} important for the simulation of multispecies 
(anti-)ferromagnetism \cite{gorshkov2010,rey2014,pagano2014,barros2017}, 
synthetic dimensions and effective quantum Hall systems \cite{mancini2015}, 
two-flavor symmetry locking \cite{lepori2015}. We note also that 
SU($m$) invariance may hold approximately also for atomic species such 
as $^{87}$Rb or $^{40}$K \cite{Giorgini08,Stamper-Kurn13}. 

The discussion above suggests that our method provides a new {\color{black} method} for efficiently probing the physics of interacting imbalanced mixtures. As a first application, we focus {\color{black}  on an attractive} two-species fermionic mixture, showing that its superfluid properties under a Rabi coupling
are the same as those of an imbalanced mixture in the absence of a Rabi term. {\color{black} This example is particularly interesting in view of the rich phenomenology of the imbalanced Fermi gases as imbalance and interaction vary, also including FFLO physics (see e.g. \cite{pethick,marchetti} and references therein).} Moreover, we argue that this mapping allows for engineering spatially and temporally dependent imbalances. They can be realized, for instance, by driving the intensity of the radio-frequency or of the Raman 
laser pulse which can induce Rabi oscillation, 
as recently done in Floquet spin-orbit coupling experiment 
\cite{jimenez-garcia14}.

The paper is organized as follows. After a general survey on the proposed method ({\color{black} Section} \ref{casoadue}), also in the presence of two-body interactions, we discuss the application of the Rabi coupling on {\color{black} an atomic mixture of two hyperfine species in a realistic experimental setup, {\color{black} and 
its generalization} to $N$-species mixtures (Section \ref{general}).}
{\color{black} As a first example,} we study a BCS/BEC superfluid in a mixture of {\color{black} two balanced hyperfine species,} coupled by a Rabi pulse (Section \ref{bcs-rabi}), finding agreement, even at the mean-field level, with the known literature on the two-species imbalanced fermionic superfluids in the absence of the Rabi coupling. In Section \ref{continuous} we reconsider the same problem in the continuous space, discussing the effect of the Rabi coupling on the renormalization of the mean-field self-consistency equations for the nontrivial order parameters. In Section \ref{applications} we address further applications, involving the possibility of tuning {\color{black} the effective population imbalances} in time and space, showing their experimental feasibility. Finally, in Section \ref{conclusions} we collect and discuss the main results, as well as possible future developments.

%%%%%%%%%%%%%%%%%%%%%%%%%%%%%%%%%%%%%%%%%%
\section{Two-species mixtures coupled by a Rabi coupling}
\label{casoadue}

In this Section we describe a general lattice setup where a mixture of  two  bosonic or fermionic {\color{black} hyperfine species} (labelled by  $\sigma = \{\uparrow \, , \downarrow \}$) 
are coupled by a Rabi term. {\color{black} When not stated otherwise, these two species are supposed to have 
the same  filling: $n_{\uparrow} = n_{\downarrow}$, with $n_{\sigma} \equiv \frac{N_{\sigma}}{V}$, $N_{\sigma}$ being the number of fermions of the species $\sigma$ on the lattice and $V$ the number of sites of the lattice.}

A Rabi term can be induced by microwave or radio-frequency techniques, or by two-photon Raman transitions induced by two opportunely detuned laser beams \cite{steck,fallanibook}. 
The latter technique also allows for a spatial dependence of the Rabi frequency, as described in Section \ref{applications}.

For simplicity, we will consider a three dimensional cubic lattice, but, as discussed in the following, no {\color{black} significant} changes arise in the continuum space or in lower dimensions. 
The atoms can hop between nearest neighbor sites with hopping energy 
$t$ ($t>0$) and  interact via a contact attractive potential with magnitude $U$.
 The Rabi coupling flips the spin-$\sigma$ orientations of the atoms, with frequency {\color{black}$\Omega(t,\b{r})$ which can be a complex value and may depend on time and position. 
We consider a real valued $\Omega(t,\b{r})$ of a factorized form in time and space, $\Omega(t,\b{r})= \Omega \,  f(t) g(\b{r})$. While by microwave or radio-frequency pulses,  a non-constant $g(\bf{r})$ is hard to 
be obtained, by back-reflecting  on a mirror the lasers inducing the Raman transitions, one can achieve a sinusoidal profile $g(\b{r}) \propto \cos ({\bf r}\cdot{q})$, where $\b{q}$ is the difference between the wave-vectors of the two lasers. }
 
 In the discussion below we omit the space and time dependences, whose consequences will be discussed at the end of the paper, in Sect. \ref{applications}.

\subsection{Fermions}
We focus first on fermions.
Thus, the Hamiltonian describing the system that we are interested in is
\beq
H = -t \sum_{<i,j>, \sigma} 
\cop{c}_{i \sigma} \aop{c}_{j \sigma} 
+ \, \Omega \sum_{i} \, 
\left( e^{i \varphi} \, \cop{c}_{i \uparrow} \, \aop{c}_{i \downarrow} + e^{-i \varphi} \, \cop{c}_{i \downarrow} \, \aop{c}_{i \uparrow} \right) - U \sum_{i} \aop{n}_{i \uparrow} \aop{n}_{i \downarrow} \, ,
 \label{H_rabi}
\eeq
where 
$ \aop{n}_{i \sigma}=\cop{c}_{i \sigma} \aop{c}_{i \sigma}$ are the number operators, {\color{black} and $\phi$ a generic phase}.

Here we describe the system in the canonical ensemble, {\color{black} as for a single experimental realization,} where the number of the atoms is fixed. 
The Hamiltonian (\ref{H_rabi}) can be mapped by the 
 unitary transformation 
 \beq
a_{i \pm} = \frac{e^{-i\frac{\varphi}{2}}c_{i \up} \pm e^{i\frac{\varphi}{2}}c_{i \down}}{\sqrt{2}} \, .
\label{rot}
\eeq
to the following Hamiltonian
\beq
H_{\mathrm{ROT}} = - t  \sum_{<i,j>, \alpha} \, \cop{a}_{i \alpha} 
\aop{a}_{j \alpha} \,  +  \Omega  \, \sum_{i}  \left( n_{i+} 
- n_{i-} 
\right)  
-  U  \sum_{i} n_{i+}n_{i-}\,,
\label{H_rot}
\eeq
where $\alpha = \pm$ and $\aop{n}_{i \pm}=a_{i \pm}^{\dag} a_{i \pm}$. 
We find that the original Rabi coupling $\propto \Omega$ is mapped to an energy imbalance $h = 2  \, \Omega$, as for a Zeeman term. Even more importantly, the interaction $\propto U $ transforms covariantly under the same transformation \cite{lepori2017}, allowing to probe the Hubbard physics also {\color{black} on imbalanced Hamiltonians (an example will be given in the next following).} 
We finally mention finally that, due to the Rabi coupling, an imbalance at $\Omega = 0$ does not produces any qualitative difference, compare to the balanced case, once $\Omega$ is switched on (see e. g. \cite{abad2013}).

The Zeeman term can be also interpreted as a chemical potential term. Indeed, as we will describe in Section \ref{bcs-rabi}, although the total number of particles $N_+ + N_-= N_\uparrow + N_\downarrow \equiv N$ remains constant {\color{black} in the present canonical scheme},  the   number $N_{\pm}$ are not fixed but can fluctuate, therefore $N_{\pm}$ can be considered as quantum averages. In particular, we will discuss the behaviour of the average difference $(N_+ - N_-)$ as a function of $U$ and $\Omega$, non vanishing in a normal state. 

The calculations above are valid also in the grand canonical ensemble, 
introducing a chemical {\color{black} potential $\mu$, then letting $N_{\uparrow}$ ($n_{i, \uparrow}$) and $N_{\downarrow}$ ($n_{i, \downarrow})$ to fluctuate, then to be fixed only in average.} In real experiments the 
$\mu$-term in (\ref{H_rabi2_mu}) can be {\color{black} ascribed} to the average occupation numbers on different experimental realizations. The resulting Hamiltonian reads:
\beq
H = -t \sum_{<i,j>, \sigma} 
\cop{c}_{i \sigma} \aop{c}_{j \sigma} 
{}\\
+ \, \Omega \sum_{i}  \left(\cop{c}_{i \uparrow} \, \aop{c}_{i \downarrow} + \cop{c}_{i \downarrow} \, \aop{c}_{i \uparrow} \right) - U \sum_{i} \aop{n}_{i \uparrow} \aop{n}_{i \downarrow} - \mu  \sum_{i, \sigma} \cop{c}_{i \sigma} \aop{c}_{i \sigma}  \, ,
 \label{H_rabi2_mu}
\eeq
and the transformation (\ref{rot}) yields:
\begin{align}
H_{\mathrm{ROT}} = - t  \sum_{<i,j>, \alpha} \, \cop{a}_{i \alpha} \aop{a}_{j \alpha} 
\,  - \, \big(\mu - \Omega \big) \, \sum_{i}  n_{i+} 
- \, \big(\mu + \Omega \big) \, \sum_{i} n_{i-}
-  U  \sum_{i} n_{i+}n_{i-} = \nonumber \\
{\color{black} = - t  \sum_{<i,j>, \alpha} \, \cop{a}_{i \alpha} \aop{a}_{j \alpha} 
\,  - \, \mu  \, \sum_{i}  \big(n_{i+} + n_{i-}\big)
+ \,  \Omega  \, \sum_{i} \big(n_{i+} - n_{i-}\big)
-  U  \sum_{i} n_{i+}n_{i-} }
\, .
\label{H_rabi_gen_2_gran_rot}
\end{align}
We find that the Rabi term in the old basis $\{ \uparrow, \downarrow \}$ became, after the transformation (\ref{rot}), an imbalance term $\delta \mu = 2 \, \Omega$ for the chemical potentials of the new species $a_{\pm}$. 

{\color{black} \subsection{Bosons}}
For a bosonic system, three Hubbard interactions are possible, intra-species interactions $U_{\uparrow\uparrow}$, 
$U_{\downarrow\downarrow}$, and an inter-species one $U_{\uparrow\downarrow}$, so that we get
\beq
H = -t \sum_{<i,j>, \sigma} 
 \cop{c}_{i \sigma} \aop{c}_{j \sigma} 
+ \, \Omega \sum_{i} \,
\left( e^{i \varphi} \, \cop{c}_{i \uparrow} \, \aop{c}_{i \downarrow} + e^{-i \varphi} \, \cop{c}_{i \downarrow} \, \aop{c}_{i \uparrow} \right) - \sum_{i}\left(
U_{\uparrow\uparrow} \,\aop{n}_{i \uparrow}^2 +
U_{\downarrow\downarrow} \, \aop{n}_{i \downarrow}^2 +
U_{\uparrow\downarrow} \, \aop{n}_{i \uparrow} \aop{n}_{i \downarrow}\right) .
 \label{H_rabi_b}
\eeq 
The interacting term of this Hamiltonian transforms covariantly under the rotation in Eq.~(\ref{rot}) iff
$U_{\uparrow \uparrow} = U_{\downarrow \downarrow} = U_{\uparrow \downarrow}/2$, that is the case for earth-alkaline atoms \cite{gorshkov2010,rey2014}. In this case, with {\color{black} these} relations on the interaction strengths, {\color{black} applying} Eq.~(\ref{rot}), Eq.(\ref{H_rabi_b}) maps to 
\beq
H_{\mathrm{ROT}} = - t  \sum_{<i,j>, \alpha} \, \cop{a}_{i \alpha}
\aop{a}_{j \alpha} \,  +  \Omega  \, \sum_{i}  \left( n_{i+}
- n_{i-}
\right)
-  \frac{U_{\up\down}}{2}  \sum_{i} (n_{i+}+n_{i-})^2\,.
\label{H_rot_b}
\eeq

{\color{black} \subsection{Trap effects}}
In current ultracold atoms experiments, both for bosons and fermions, a typical ingredient is the external trapping potential {\color{black} of the form
\beq
\sum_i V(\b{r}_i) \, ( \aop{n}_{i \uparrow} + \aop{n}_{i \downarrow}) \, ,
 \label{trap}
 \eeq 
where $V(\b{r}_i)  = \frac{m}{2} \omega^2 |\b{r}_i|^2$ and} $\b{r}_i$ is the vector distance of the lattice site at position $i$ from the center of the trap. We assume the trapping frequency $\omega$ and the mass $m$ equal for both the species $\{\uparrow,\downarrow\}$, as in most of the experiments involving different hyperfine levels of the same atom. 
  It is straightforward to show that also the potential in  (\ref{trap}) transforms covariantly under the rotation in  Eq.~(\ref{rot}):
{\color{black}
\beq
\sum_i  V(\b{r}_i)  \, ( \aop{n}_{i +} + \aop{n}_{i -}) \, , 
 \label{trap2}
 \eeq
} 
then all the previous discussions are not spoiled by the presence of this term.

In the light of the above consideration, we expect that the properties of balanced mixtures (both in the canonical and in the {\color{black} grandcanonical} ensemble) under the Rabi coupling and possibly {\color{black} of two-body interactions} (provided that the 
interactions do not to spoil the rotated states  (\ref{rot})) to be  
equal to the physics of imbalanced interacting mixtures. This 
will be {\color{black} further studied} in Section (\ref{bcs-rabi}), considering the superfluid properties of the Hamiltonian (\ref{H_rot}).\\

\subsection{Comments on the experimental implementation}
\label{exp}
In this subsection, we comment on the limits of tunability of the 
Rabi coupling, also in relation with the other parameters,
as the typical hopping amplitude $\{t\}$ and the strengths of the interaction 
$\{U_i\}$. Indeed our method, to be effective, requires
the Rabi coupling to be tunable over a range of energies at least 
comparable with the smallest energy scales between the hopping amplitude and the interactions. 
This condition is easily fulfilled for constant (or smoothly varying) $\Omega$. 
Indeed, it is possible to achieve $\Omega \sim t$ within the validity 
of the tight-binding approximation. 
This favourable  situation can change if the Rabi frequency depends on the position or on the time. Roughly speaking, the allowed dependence should fulfill 
the adiabatic theorem. 
In Sect. \ref{modulated}, we quantitatively discuss the limits imposed 
by the presence of a sinusoidal behavior for $\Omega$ induced by Raman lasers. 
In particular
we will show that the related momentum transfer does not considerably limit further the range of allowed values for $\Omega$. 
Thus, it turns out that there are neither conceptual nor technical limitations in the use of Rabi coupling technique. Nowadays intensities 
$\Omega \sim$ kHz are realistic, which are of the same 
order of magnitude of the Fermi energies in typical experiments, 
both in the continuous space and on the lattice. 
We conclude that unbalancing an attractive two-species {\color{black} 
hyperfine mixture} by a Rabi coupling allows to investigate (at least) 
both the BCS and BEC limits of the superfluid regime for ultracold 
fermionic mixtures.

{\color{black} \subsection{Further generalizations}}
In principle, a spin-dependent hopping $t_{\sigma}$ can be added to  (\ref{H_rabi2_mu}), {\color{black} for instance exploiting superlattice configurations 
(see, e.g., \cite{lewenstein12} and references therein).} 
Interestingly, if we apply the transformation (\ref{rot}) to the resulting Hamiltonian, we obtain:
\begin{eqnarray}
\nonumber 
H_{t \, , \, \mathrm{ROT}} &=& - \frac{t_{\uparrow} + t_{\downarrow}}{2} \sum_{<i,j>, \alpha}   \cop{a}_{i \alpha} \aop{a}_{j \alpha} 
                        -  \frac{t_{\uparrow} - t_{\downarrow}}{2}  \sum_{<i,j>}  \big( \cop{a}_{i +} \aop{a}_{j -}  + \cop{a}_{i -} \aop{a}_{j +}  \big) 
\\
&&
- \, \big(\mu - \Omega \big) \, \sum_{i}  n_{i+}  
-  \, \big(\mu + \Omega \big) \, \sum_{i} \aop{n}_{i-}        
-  U  \sum_{i} n_{i+} n_{i-}\, ,
\label{H_rabi_gen_4_rot}
\end{eqnarray}
We find that, in the rotated frame, in addition to the spin-dependent chemical potential, a spin-orbit-like coupling term appears. {\color{black} Therefore, this scheme can be used to simulate a spin-orbit coupling, at least in one-dimensional lattices.}
If we proceed further with the diagonalization, we end up with the Hamiltonian 
\beq
\begin{array}{c}
H_{t \, , \,  \mathrm{ROT}}^{(\mathrm{diag})} = \sum_{\b{k}} \Big[ \lambda_+(\b{k}) \, \cop{a}_{\bf{k} +} \aop{a}_{\b{k} +}  
+   \lambda_-(\b{k}) \, \cop{a}_{\b{k} -} \aop{a}_{\b{k} -} -  U  n_{\b{k}+} n_{\b{k}-} \Big]\,, 
\end{array}
\label{H_rot_sbil_rot_corr}
\eeq
with
\beq
\lambda_{\pm}(k) = {\varepsilon}_\b{k} - \mu \pm \sqrt{\delta \varepsilon_\b{k}^2 + \Omega^2}  \, ,
\eeq
where ${\varepsilon}_\b{k} \equiv \frac{\varepsilon_{\b{k}\uparrow} + 
\epsilon_{\b{k}\downarrow}}{2}$  
and $\delta \varepsilon_\b{k} \equiv \frac{\varepsilon_{\b{k}\uparrow} - \varepsilon_{\b{k}\downarrow}}{2}$, posing  
$\varepsilon_{\b{k}\uparrow}=-t_\uparrow\sum_{s=x,y,z}\cos k_s$,
$\varepsilon_{\b{k}\downarrow}=-t_\downarrow\sum_{s=x,y,z}\cos k_s$.

%%%%%%%%%%%%%%%%%%%%%%%%%%%%%%%%%%%%%%%%%%
\section{Generalization to $N$ species}

\label{general}
The model described in the previous Section can be easily extended to a 
$N-$mixture ($N = 2 M$ even) of bosonic or fermionic atoms in several ways, 
for instance by coupling the various {\color{black} hyperfine species} pairwise.
The non-interacting  grand canonical Hamiltonian with possible imbalances in the densities (with different chemical potentials) and in the hopping amplitudes reads in this case:
\bea
\nonumber 
H_{\{\mu_l\} ,  \{t_l\}} &=& - \sum_{<i,j>, l}  \, t_{2l}  
\cop{c}_{i  , 2l} \aop{c}_{j , 2l} 
- \sum_{<i,j>, l}  \, t_{2l+1} 
\cop{c}_{i  , 2l+1} \aop{c}_{j , 2l+1} 
\\
&&-  \sum_{i, l} \, \big(\mu_{2l} \, \cop{c}_{i \, 2l} \aop{c}_{i , 2l} +  \mu_{2l+1} \, \cop{c}_{i \, 2l+1} \aop{c}_{i , 2l+1} \big)  + \, \sum_{i, l} \, {\Omega_l} \,  \left( \cop{c}_{i , 2l} \, \aop{c}_{i , 2l+1} +  \cop{c}_{i , 2l+1} \, \aop{c}_{i , 2l} \right) ,
 \label{H_rabi_gen_5}
\eea
where $l =1, \dots, M$ labels the pairs of {\color{black} hyperfine species} coupled by the Rabi coupling.

This Hamiltonian, after the unitary transformation 
\beq
a_{i \pm}^{(l)} = \frac{c_{i , 2l} \pm c_{i , 2l+1}}{\sqrt{2}} \, ,
\label{rot_gen}
\eeq
becomes
\bea
&&H_{\{\mu_l\} , \{t_l\} \, \mathrm{ROT}} = - \sum_{<i,j>, \alpha , l}  \frac{t_{2l} + t_{2l+1}}{2} \, {a}_{i \alpha}^{(l)\dag} a_{j \alpha}^{(l)} 
\,  
- \sum_{<i,j>, l} \frac{t_{2l} - t_{2l+1}}{2}  \left( a_{i +}^{(l)\dag} a_{j -}^{(l)}  + a_{i -}^{(l)\dag} a_{j +}^{(l)}  \right)\\
\nonumber &&
- \sum_{i,l}\left[  \left( \frac{\mu_{2l} + \mu_{2l+1}}{2} - \Omega_l \right)  
 n^{(l)}_{i+} 
+ \left( \frac{\mu_{2l} + \mu_{2l+1}}{2} + \Omega_l \right)  
n^{(l)}_{i-} \right]  
- \sum_{i, l}  \frac{\mu_{2l} - \mu_{2l+1}}{2}  
\left( a_{i +}^{(l)\dag} a_{i -}^{(l)}  + a_{i -}^{(l)\dag} a_{i +}^{(l)}  \right).  
\label{H_rot_gen_5}
\eea
Possible density-density interactions have the same form also in the dressed basis (\ref{rot_gen}), provided that the interactions involve only the {\color{black} hyperfine species} pairwise $2l$-$(2l+1)$:
\beq
\sum_{l=1}^M \sum_{i,j} \, V_l (i-j) \, \Big( n_{i , 2l} + n_{i , 2l+1} \Big) \Big( n_{j , 2l} + n_{j , 2l+1} \Big)
\eeq
($n_{i , \alpha} = c^{\dag}_{i,\alpha} c_{i,\alpha}$).
In this case, the transformation (\ref{rot_gen}) yields 
\beq
\sum_{l=1}^M \sum_{i,j} \, V_l (i-j) \,  \Big( n^{(l)}_{i +} + n^{(l)}_{i  -} \Big) \Big( n^{(l)}_{j +} + n^{(l)}_{j -} \Big) \, ,
\eeq
where $n^{(l)}_{i  \pm}=a^{\dagger (l)}_{i  \pm}a^{(l)}_{i  \pm}$. 

Another case, {\color{black} even more interesting from the experimental point of view}, occurs when the interactions pair all the {\color{black} hyperfine species} in the mixtures, with the same strength  and the same space dependence {\color{black} (as for earth-alkaline-like atoms \cite{gorshkov2010,rey2014,pagano2014,mancini2015}, see also the Introduction):}
\beq
 \sum_{i,j} \, V (i-j) \, \sum_{s, s^{\prime} = 1}^{2M} \,  n_{i , s} \,  n_{j , s^{\prime}} =  \sum_{i,j} \, V (i-j) \, \Bigg(\sum_{s=1}^{2M} \,  n_{i , s} \Bigg) \,  \Bigg(\sum_{s^{\prime} = 1}^{2M} \,  n_{j , s^{\prime}} \Bigg) \, .
\eeq
This interaction still transforms covariantly under \eqref{rot_gen}:
\beq
 \sum_{i,j} \, V (i-j) \, \, \sum_{l , l^{\prime} = 1}^{M} \, \Big(n^{(l)}_{i  +} + n^{(l)}_{i  -}\Big) \Big(n^{(l^{\prime})}_{j  +} + n^{(l^{\prime})}_{j  -} \Big) \, .
\eeq
In these cases, the dynamics can be described easily in terms of dressed states as well, since the scattering processes related to the interactions do not spoil them. 

A clarification of the latter point can be obtained considering the example of two species, $\up$ and $\down$ (labelled by the momentum),
scattering in the $s$-wave channel. 
 Due to the Pauli principle, the low-energy scattering, supposed elastic, can occur only in the channel 
 \beq
\frac{1}{\sqrt{2}} \left[ \frac{\up + \down}{\sqrt{2}} \otimes \frac{\up - \down}{\sqrt{2}} - \frac{\up - \down}{\sqrt{2}} \otimes \frac{\up + \down}{\sqrt{2}} \right]  (\b{k}) = \frac{\up \down - \down \up}{\sqrt{2}}  (\b{k}) \to  e^{i \phi_{\b{k}}} \, \frac{\up \down - \down \up}{\sqrt{2}}  (\b{k}) \, ,
\label{scat}
\eeq 
$\phi_{\b{k}}$ being the momentum depending scattering phase.  
We find that $e^{i \phi_{\b{k}}}$ globally multiplies the final scattering state, without spoiling the relative coherence of the states $\up \down$ and $\down \up$. 
This is an effect of the linearity of the scattering matrix \cite{landauqm} and it immediately holds if $N=2$. 
The phase $e^{i \phi_{\b{k}}}$ is nothing but the phase resulting from the scattering at $\Omega = 0$ of $\up$ and $\down$ particles, with momentum $\b{k}$ in the spin singlet state. 

In the general case, with $N \ge 2 $, the conditions that we imposed above on the interactions assure that every scattering process changes the initial state $\frac{|l\rangle \pm |l+1\rangle}{\sqrt{2}}$ just 
by multiplying it by a pure phase $e^{i \phi_{l\b{k}}}$, as for the case reported in Eq.~(\ref{scat}).

%%%%%%%%%%%%%%%%%%%%%%%%%%%%%%%%%%%%%%%%%%
\section{Two-species superfluidity in the presence of a Rabi coupling}
\label{bcs-rabi}

In this Section, we exemplify {\color{black} the effect of an imbalance induced by a Rabi coupling}, focusing  on a two-species fermionic mixture and studying the properties of its superfluid phase. We consider the Hamiltonian (\ref{H_rabi2_mu}): 
\beq
H = -t \sum_{<i,j>, \sigma}
\cop{c}_{i \sigma} \aop{c}_{j \sigma} 
- \mu  \sum_{i, \sigma} \cop{c}_{i \sigma} \aop{c}_{i \sigma}
+ \, \Omega \sum_{i}  \left(\cop{c}_{i \uparrow} \, \aop{c}_{i \downarrow} + \cop{c}_{i \downarrow} \, \aop{c}_{i \uparrow} \right) - U \sum_{i} \aop{n}_{i \uparrow} \aop{n}_{i \downarrow} \, .
\label{H_target}
\eeq
As discussed in Section \ref{casoadue}, we explicitly include the chemical 
potential term. 
{\color{black} In the following, this Hamiltonian and similar ones having an Hubbard interaction, will be denoted as "full" Hamiltonians, 
in contrast to the mean field quadratic Hamiltonians}.

 In the presence of a nonzero superfluid gap $\Delta$ {\color{black} (to be verified a posteriori)}, the corresponding {\color{black} (mean field)} BCS-projected Hamiltonian is:
 \begin{equation}
 H= -t \sum_{<i,j>, \sigma} 
\cop{c}_{i \sigma} \aop{c}_{j \sigma} 
- \left( \mu + \frac{Un}{2} \right) \sum_{i, \sigma} \cop{c}_{i \sigma} \aop{c}_{i \sigma} +
 \left( \Omega + \gamma \right)  \sum_{i} \left( \cop{c}_{i \uparrow} \aop{c}_{i \downarrow} + \cop{c}_{i \downarrow} \aop{c}_{i \uparrow} \right) -\Delta \sum_{i} \left( \cop{c}_{i \uparrow} \cop{c}_{i \downarrow} + \aop{c}_{i \downarrow} \aop{c}_{i \uparrow}  \right) \, .
\label{Hproj_rabi}
\end{equation}
In Eq.~(\ref{Hproj_rabi}) we assumed the presence of a  {\color{black} further bilinear order parameter, also coming from the Wick decomposition of the Hubbard interaction term in Eq.  \eqref{H_target}}
\beq
\gamma= -U \langle  \cop{c}_{i \uparrow} \aop{c}_{i \downarrow} \rangle = - U \langle \cop{c}_{i \downarrow} \aop{c}_{i \uparrow}\rangle \, ,
\label{gammadef}
\eeq
due to the presence of the Rabi coupling. Moreover, we fix $(\Omega + \gamma) >0$, up to a phase redefinition of the ${c}_{i \sigma}$ operators. 
Going to momentum space, the Hamiltonian
(\ref{Hproj_rabi}) can be easily diagonalized, finding the eigenvalues: 
\beq
\lambda_{\b{k}}^{(\pm)} = E_{\b{k}} \pm (\Omega +\gamma)  \, ,
\label{spectrum}
\eeq
where $E_{\b{k}} = \sqrt{\xi_{\b{k}}^2 + \Delta^2}$ and
\beq
\xi_{\b{k}}= \varepsilon_{\b{k}} - \tilde{\mu} \, , \, \quad {}  \quad \,
\varepsilon_{\b{k}}= -2t \sum_{l= x,y,z} \cos k_l \, ,  \, \quad {}  \quad \, 
\tilde{\mu}=\mu + \frac{UN}{2V} \, .
\label{xi}
\eeq
The Bogoliubov coefficients turn out to be the same as in the purely ($\Omega = 0$) BCS case. However, we find that, introducing the Rabi term, the quasiparticle excitation spectrum is split into two bands, shifted by $\pm(\Omega + \gamma)$ with respect to the usual BCS case.
Correspondingly, the minimum of the excitation spectrum is 
\begin{align}
\Delta_G=\Delta - \left| \Omega + \gamma \right| \, .
\end{align}
The ground-state energy reads:
\beq
E_{\mathrm{SUP}} =  \frac{V \, \Delta^2}{U}  + \frac{1}{2} \sum_{\b{k}} \Big[ \left( \xi_{\b{k}} - \lambda^{(+)} (\b{k}) \right) + \left( \xi_{\b{k}} -  \lambda^{(-)} (\b{k}) \right) \Big] +  \sum_{\b{k} \in \bar{\mathcal{D}}} \lambda^{(-)} (\b{k})   \, ,
\label{egs}
\eeq
where $\mathcal{D}$ is a  domain in the first Brillouin zone defined as
\begin{align} \label{D}
\mathcal{D}= \left\{ \b{k} \in 1^{\mathrm{st}} \mathrm{BZ} :  E_{\b{k}} > \Omega + \gamma \right\}
\end{align} 
and its complementary as 
\begin{align} \label{CD}
\bar{\mathcal{D}}= \left\{ \b{k} \in 1^{\mathrm{st}} \mathrm{BZ} :  E_{\b{k}} < \Omega + \gamma \right\} \, .
\end{align}
The explicit calculation of the self-consistent equations for the order parameters $\Delta, \gamma, \mu$ is performed by setting to 0 the derivatives of $E_{\mathrm{SUP}} + \mu N+ \gamma \, \delta N$, {\color{black} $\delta N \equiv (N_{-} - N_{+})$,}  with respect for $\Delta$, $\gamma$ and $\mu$.
 {\color{black} The introduction of the quantity $\gamma \, \delta N$ stems from the Wick decomposition of the Hubbard interaction in \eqref{H_target}, similar as the first term in  $E_{\mathrm{SUP}}$ \cite{annett}}.  The final result is:
\begin{align}
& 1 = \frac{U}{2V} \sum_{\b{k} \in \mathcal{D}} \, \frac{1}{E_{\b{k}}} \, ,\label{sc_delta1}\\
& \delta N = -  \sum_{\b{k} \in \bar{\mathcal{D}}} \, 1 \, ,  \label{sc_gamma1}\\
&n=  \frac{1}{V} \, \sum_{\b{k} } \, \Big( 1- \frac{\xi_{\b{k}}}{E_{\b{k}}} \Big) + \frac{1}{V} \, \sum_{\b{k} \in  \bar{\mathcal{D}}} \, \frac{\xi_{\b{k}}}{E_{\b{k}}}    = 1- \frac{1}{V} \, \sum_{\b{k} \in \mathcal{D}} \, \frac{\xi_{\b{k}}}{E_{\b{k}}}  \, .
 \label{sc_mu1}
\end{align}
From (\ref{sc_gamma1}), we conclude that $\gamma \leq 0$, with $\gamma = 0$ (as well as $\bar{\mathcal{D}} =0$) iff $\Omega = 0$; moreover in (\ref{sc_mu1}) an additional term is present, compared to the $\Omega = 0$ case.

We observe that the domain $\mathcal{D}$, as well as its complementary  $\bar{\mathcal{D}}$, differs from the definition given implicitly in \cite{marchetti}, 
where  the contribution of the imbalance $\delta N$  has been neglected in the Hartree terms $- U N_- \sum_i a_{i +}^{\dagger} a_{i +} - U N_+ \sum_i a_{i -}^{\dagger} a_{i -}$. In this case, it holds $\mathcal{D}= \left\{ \b{k} \in 1^{\mathrm{st}} \mathrm{BZ} :  E_{\b{k}} > \Omega + \gamma \right\}$ and 
the equation (\ref{sc_gamma1}) is simplified. This aspect will be discussed in Section \ref{continuous}, where we will infer  that, if $U=0$, the parameter {\color{black} $\delta N$} equals the imbalance of two fermionic species with chemical potentials $\mu_{\pm} \equiv \mu \pm \Omega$. The same identification holds for the ground-state energy (\ref{egs}).
 
{\color{black} \subsection{Comparison at mean-field level}}

{\color{black} The mean field ground-state energy (\ref{egs}),  resulting from the full Hamiltonian \eqref{H_target}, equals the same quantity for a imbalanced mixture under an onsite attraction \cite{marchetti}, described by the  Hamiltonian (\ref{H_rabi_gen_2_gran_rot}). Indeed, we found that the Hamiltonian (\ref{H_target}) maps exactly to (\ref{H_rabi_gen_2_gran_rot}) under the transformation (\ref{rot}); for this reason the spectra and the phenomenologies at these two full Hamiltonians must be the same. However, one can ask whether the equivalence 
exactly established is valid also at mean-field level.

In the following we show, as it may be expected, that the equivalence is valid 
as well for the 
corresponding Hamiltonians obtained in mean-field approximations, i.e., 
that the previous result holds at tme mean-field level (where the 
self-consistency conditions have to be enforced).} 

The mean-field Hamiltonian from (\ref{H_rabi_gen_2_gran_rot}) reads:
\begin{multline}
H_{\mathrm{ROT}} = - t  \sum_{<i,j>, \alpha} \, a_{i \alpha}^{\dag} a_{j \alpha} 
\,  - \, \big(\mu - \Omega \big) \, \sum_{i}  \cop{a}_{i +} a_{i +}  
-  \, \big(\mu + \Omega  \big) \, \sum_{i} \cop{a}_{i -} a_{i -} - \\
- U N_- \sum_i a_{i +}^{\dagger} a_{i +} - U N_+ \sum_i \cop{a}_{i -} a_{i -} + \Delta \sum_{i} \big(a_{i +}  a_{i -} + \mathrm{h. c.}\big)  + \mu \, N + \Theta \, \delta N\,  ,
\label{H_rabi_gen_2_gran_rot_MF}
\end{multline} 
Similarly as the previous subsection, the self-consistent equations for $\Delta$, $N$,  $\delta \, N$ can be found  deriving $E_{\mathrm{SUP}} + \mu N+ \Omega \, \delta N$  with respect of $\Delta$, $\Theta$ and $\mu$ \cite{marchetti}. The results {\color{black} are the same as in 
Eqs. (\ref{sc_delta1})- (\ref{sc_mu1}).} 

{\color{black} We can check at this point that the mean-field Hamiltonian 
obtained from (\ref{Hproj_rabi}) by the transformation
(\ref{rot}) 
\begin{multline}
H_{\mathrm{ROT}} = - t  \sum_{<i,j>, \alpha} \, \cop{a}_{i \alpha} a_{j \alpha} 
\,  - \, \big(\mu - \Omega  - \gamma \big) \, \sum_{i} \cop{a}_{i +} a_{i +}  
-  \, \big(\mu + \Omega +\gamma \big) \, \sum_{i} \cop{a}_{i -} a_{i -} \, -\\
 - \frac{U}{2} N \sum_i \big( \cop{a}_{i +} a_{i +} +  \cop{a}_{i -} a_{i -}\big) +
\Delta \sum_{i} \big(a_{i +}  a_{i -} + \mathrm{h. c.}\big)  + \mu \, N + (\Omega + \gamma) \, \delta N\,  ,
\label{H_rabi_gen_2_gran_rot_MF_1}
\end{multline}
coincides with the mean-field Hamiltonian (\ref{H_rabi_gen_2_gran_rot_MF}). 
Interestingly, this comparison sheds light on the physical meaning 
of the parameter $\gamma$, yielding indeed:
\beq
\gamma = \frac{U}{2} \, \delta N \, .
\label{relazione}
\eeq
Summing up, we found that the relation \eqref{relazione} for the mean-field parameter $\gamma$ is imposed by the request of equality between the mean-field Hamiltonians \eqref{H_rabi_gen_2_gran_rot_MF} and \eqref{H_rabi_gen_2_gran_rot_MF_1}, a fact physically motivated, but not trivially implied, by the equivalence of the full Hamiltonians \eqref{H_target} and \eqref{H_rabi_gen_2_gran_rot}.}\\

\subsection{Discussion}
Since the Hamiltonians (\ref{H_rabi_gen_2_gran_rot}) and (\ref{H_rabi_gen_2_gran_rot_MF}) have been studied extensively in various papers devoted to imbalanced fermionic mixtures \cite{pethick,pao}, both at zero and at finite  temperatures, we do not further study their properties, and refer to the pertinent 
literature. For our purposes, indeed, the main point we want to stress here is the equivalence of these Hamiltonians in the presence of Rabi couplings with those describing imbalanced Fermi mixtures. {\color{black} In the experiments, 
one performs measurements on the fermionic species $c$'s; from 
the obtained findings for the quantities $\langle c^\dag c \rangle$, $\langle c^\dag c^\dag \rangle$, and via the relation between the $c$'s and 
the $a$'s, one can then reconstruct the phase diagram of the 
($a$-)imbalanced mixture.}
The resulting main feature, based on the available results, is the appearance, both on the BCS and BEC sides,  of a coexistence of normal and superfluid phases, for suitable effective imbalances $\delta \mu = 2 \Omega$. Moreover, unconventional forms of superconductivity, as FFLO, are conjectured, for an extended review, see \cite{SR}.  
By the definition of $\gamma$, we expect that $\gamma = 0$ in a superfluid phase, where $N_+ = N_-$.

%%%%%%%%%%%%%%%%%%%%%%%%%%%%%%%%%%%%%%%%%%

\section{The continuous case}
\label{continuous}

In this Section we deal with the analogous of (\ref{H_rabi}) in the continuum space.  In this situation, most of the features are qualitatively equal to 
the lattice case, then we detail {\color{black} only 
the formulation and discussion of the equations} for  $\gamma$, $\mu$ and $\Delta$. In particular we discuss the removal of the infinities encountered during their solution. 
We assume again to work in three dimensions. 
The Hamiltonian is:
\bea
H^{(\mathrm{cont})} &=&\frac{1}{V} \, \sum_{\sigma} \, \int \mathrm{d} \b{r} \ \cop{c}_{\sigma} (\b{r})  \Big( - \hbar^2 \,  \frac{\nabla^2}{2m} - \mu  \Big) \, \aop{c}_{\sigma}  (\b{r})   \, +  \nonumber \\ 
&&+\, \frac{\Omega}{V} \, \int \mathrm{d} \b{r} \, \left( \cop{c}_{\uparrow}  (\b{r}) \aop{c}_{\downarrow}  (\b{r}) + \cop{c}_{\downarrow}  (\b{r}) \aop{c}_{\uparrow}  (\b{r}) \right) - \frac{U}{V} \, \int \mathrm{d} \b{r} \, \aop{n}_{\uparrow} (\b{r}) \,  \aop{n}_{\downarrow} (\b{r}) \, ,
\label{H_cont}
\eea
$V$ denoting  the volume in this Section.
The eigenstates of (\ref{H_cont}) in the absence of the interaction term $H_i = -  \frac{U}{V} \, \int \mathrm{d} \b{r} \, \aop{n}_{\uparrow} (\b{r}) \,  \aop{n}_{\downarrow} (\b{r})$ are superpositions of plane waves with momentum $\b{k}$, of the form
$a_{\pm}(\b{k})=\frac{c_{\up} \pm c_{\down}}{\sqrt{2}} (\b{k})$, as in (\ref{rot}). 
These states are assumed to be asymptotical in the scattering evolution \cite{Giorgini08}, and the interaction
to arise between them without coherence spoiling effects. Since here $N=2$ in the notation of Section \ref{general}, this requirement is automatically fulfilled (see the discussion therein).
The Hamiltonian transformed by (\ref{rot}) reads:
\bea
H^{(\mathrm{cont})}_{\mathrm{ROT}} &=& - \frac{\hbar^2}{V} \, \sum_{\alpha = \pm} \, \int \mathrm{d} \b{r} \ \cop{a}_{\alpha} (\b{r})  \, \frac{\nabla^2}{2m} \, \aop{a}_{\alpha}  (\b{r})  \,  -
  \frac{\big( \mu - \Omega \big)}{V} \,
 \int \mathrm{d} \b{r} \,  \cop{a}_{+}  (\b{r}) \aop{a}_{+}  (\b{r}) \nonumber \\
&&- \frac{\big( \mu + \Omega \big)}{V} \, \cop{a}_{-}  (\b{r}) \aop{a}_{-}  (\b{r})  - \frac{U}{V} \, \int \mathrm{d} \b{r} \, \aop{n}_{+} (\b{r}) \,  \aop{n}_{-} (\b{r}) \, .
\label{H_cont2}
\eea

\subsection{Gap equations}
We consider the zero-temperature self-consistency equations for 
$\gamma$, $\mu$, and $\Delta$.  
These equations have the same functional form as in the lattice case: 
\begin{align}
& 1 = \frac{U}{2} \, \frac{V}{(2 \pi \hbar)^3} \int_{\mathcal{D}} \mathrm{d} \b{k} \,  \frac{1}{E_{\b{k}}} \label{sc_delta_cont}\\
&\gamma = - \frac{U}{2} \, \frac{V}{(2 \pi \hbar)^3} \int_{\mathcal{\bar{D}}} \mathrm{d} \b{k} \,  1 \label{sc_gamma_cont}\\
&n=  1 - \frac{1}{(2 \pi \hbar)^3} \int_{\mathcal{D}} \mathrm{d} \b{k} \,  \frac{\xi_{\b{k}}}{E_{\b{k}}}  \, ,
 \label{sc_mu_cont}
\end{align}
similarly, the domains  $\mathcal{D}$ and $\mathcal{\bar{D}}$ are the same as in  Section \ref{bcs-rabi}.  Indeed, no assumption has been {\color{black} made} about the precise form of $\varepsilon_\b{k}$ in the derivation of the equations for  $\gamma$, $\mu$ and $\Delta$. Clearly, $\mathcal{D}$ and $\mathcal{\bar{D}}$ now take values  into the infinite set of all the possible three dimensional momenta, and not any longer in the first Brillouin zone. 
In order to derive (\ref{sc_delta_cont})(\ref{sc_mu_cont})(\ref{sc_gamma_cont}) we also used the fact that $\sum_{\b{k}} \to \frac{V}{(2 \pi \hbar)^3} \, \int  \mathrm{d} \b{k}$ passing to the continuum limit.

In the absence of Rabi coupling, equation (\ref{sc_delta_cont}) is known to be divergent  and to need regularization, by the introduction of the scattering lengths $a$ \cite{pethick}.
 Let us study what happens to (\ref{sc_delta_cont}) (\ref{sc_gamma_cont}) and (\ref{sc_mu_cont}) in the presence of a Rabi term.
To do this, {\color{black} it is} useful to reconsider their derivation in the presence of a finite range potential $U(\b{r})$, generalizing the equation in Section \ref{bcs-rabi} (where  a $\delta(\b{r})$ interaction has been assumed instead). The result is:
\begin{align}
&\Delta_{\b{k^{\prime}}} = \, \frac{1}{2} \, \frac{1}{(2 \pi \hbar)^3} \int_{\mathcal{D}} \mathrm{d} \b{k} \, 
\frac{\tilde{U}(\b{k} ,\b{k^\prime})}{\sqrt{\xi_{\b{k}}^2 + \Delta_{\b{k}}^2 }} \, \Delta_{\b{k}} \, , \label{sc_finite_delta} \\
&\gamma_{\b{k^{\prime}}} = -  \, \frac{1}{2} \,  \frac{1}{(2 \pi \hbar)^3} \int_{\mathcal{\bar{D}}} \mathrm{d} \b{k} \, 
\tilde{U}(\b{k} ,\b{k^\prime}) \, , \label{sc_finite_gamma} \\
&n= 1 - \frac{1}{(2 \pi \hbar)^3} \int_{\mathcal{D}} \mathrm{d} \b{k} \,  \frac{\xi_{\b{k}}}{E_{\b{k}}}\, , \label{sc_finite_mu} 
\end{align}
where $\tilde{U} ({\b{k}, \b{k^\prime}}) = \tilde{U} (\b{k} - \b{k^\prime})$ denotes the Fourier transform of $U(\b{r})$, 
$\tilde U(\b{q})= \int \mathrm{d} \b{r} \, e^{i \b{q}\cdot \b{r}} \, U(\b{r})$. 
We insert in these equations the scattering amplitude
(for the moment not restricted to the $l=0$ contribute) $f(\b{k} , \b{k^\prime})$, via the formula \cite{landauqm, cooperp}:
\beq
f(\b{k} , \b{k^\prime}) = \frac{m}{4 \pi \hbar^2}  \, \tilde{U}(\b{k} , \b{k^\prime}) + \frac{1}{(2 \pi \hbar)^3}\int \mathrm{d} \b{q} \, \frac{\tilde{U}(\b{k^{\prime}} , \b{q}) f(\b{k}, \b{q})}{\big(\varepsilon_\b{k} -\varepsilon_\b{q}-i 0^{+}\big)}  \, .
\label{amplitude}
\eeq
The solution of this equation can be obtained by iteration on $f(\b{p} , \b{p^\prime})$ \cite{landauqm} in the second term to the right side:
\beq
 f(\b{k} , \b{k^\prime}) = \frac{m}{4 \pi \hbar^2}  \Bigg[ \tilde{U}(\b{k} , \b{k^\prime}) +  \frac{1}{(2 \pi \hbar)^3}\int \mathrm{d} \b{q} \, \frac{\tilde{U}(\b{k^{\prime}} , \b{q}) \, \tilde{U}(\b{k}, \b{q})}{\big(\varepsilon_\b{k} -\varepsilon_\b{q}-i 0^{+}\big)} + \dots \Bigg] \, ,
 \label{sol}
\eeq
the approximation keeping only the first term of (\ref{sol}) quoted as (first) Born approximation \cite{landauqm}.
 Multiplying now both of the terms in (\ref{amplitude}) for $\mathrm{d} \b{k^{\prime}}$, performing this integration, and exploiting (\ref{amplitude}), we obtain:
 \beq
\Delta_{\b{k^{\prime}}} =  \frac{1}{2 }  \, \frac{4 \pi \hbar^2}{m} \ \frac{1}{(2 \pi \hbar)^3} \Bigg( 
\int_{\mathcal{D}} \mathrm{d} \b{k} \,  \frac{ f(\b{k} ,\b{k^\prime}) }{\sqrt{\xi_{\b{k}}^2 + \Delta_{\b{k}}^2 }} - \int \mathrm{d} \b{k} \frac{f(\b{k} ,\b{k^\prime})}{\varepsilon_\b{k} - \varepsilon_\b{k^{\prime}}}  \Bigg)  \Delta_{\b{k}} \, .
 \label{sc_finite_delta_2}
\eeq
Let us restrict now the scattering in the low-energy  limit, where $\b{k}, \b{k^\prime} \to 0$: in this regime $\Delta_{\b{k}, \b{k^{\prime}}} \to \Delta$ and $ f(\b{k} , \b{k^\prime}) \to  -a$ 
($ a$ being the scattering length in $s$-wave, supposed negative) \cite{landauqm}, so that we obtain:
\beq
1 = - \frac{1}{2} \frac{4 \pi \hbar^2 a}{m} \, \frac{1}{(2 \pi \hbar)^3} \Bigg( \int_{\mathcal{D}} \mathrm{d} \b{k}  \,
 \frac{1 }{\sqrt{\xi_{\b{k}}^2 + \Delta_{\b{k}}^2 }} - \int \mathrm{d} \b{k}  \, \frac{1}{\varepsilon_\b{k}} \Bigg) \, .
 \label{sc_finite_delta_fin}
\eeq
This equation is the renormalized version of (\ref{sc_delta_cont}). Notice that the integration on the second term in the right part of (\ref{sc_finite_delta_fin}) is onto all the possible momenta, independent on the range of integration $\mathcal{D}$ of the first term, as in the absence of the Rabi term. The result is expected in the light of the mapping of (\ref{H_rabi2_mu}) on (\ref{H_rabi_gen_2_gran_rot}): the effective imbalance in the last Hamiltonian does not affect the interaction.

We deal now with equation (\ref{sc_finite_gamma}): this is finite since $\bar{\mathcal{D}}$ is finite. Using the Born approximation
$f(\b{k} , \b{k^\prime})  \approx \frac{m}{4 \pi \hbar^2} \tilde{U}(\b{k} , \b{k^\prime})$, one gets:
\beq
\gamma = - \frac{2 \pi \hbar^2 a}{m} \, \frac{1}{(2 \pi \hbar)^3} \int_{\mathcal{\bar{D}}} \mathrm{d} \b{k} \, . 
\label{gamma_cont}
\eeq
We stress that this approximation, well motivated here, is instead source of divergencies if used in the {\color{black} equations} (\ref{sc_delta_cont}) and (\ref{sc_finite_delta}). Different approximations in (\ref{sc_delta_cont}) and (\ref{sc_gamma_cont}) for the scattering amplitudes are possible since the two equations are decoupled: the latter one can be solved once the former one and (\ref{sc_mu_cont}) have been solved at the same time (see at the end of this subsection). Finally we observe that equation (\ref{sc_mu_cont}) is finite and independent on  $\tilde{U} ({\b{k} , \b{k^\prime}})$, then it does not require any further handling before to be solved.

We consider now the explicit solution of  (\ref{sc_finite_delta_fin}), (\ref{gamma_cont}) and (\ref{sc_mu_cont}): we already showed that $\Delta$ and $\gamma$ cannot be nonvanishing at the same time, for this reason we consider separately the superconductive ($\gamma = 0$, $\Delta \neq 0$) and  the normal state  regimes ($\gamma \neq 0$, $\Delta = 0$). Notice that in the case of coexistence of superfluid and normal state, as for suitable imbalanced fermionic mixtures (see \cite{SR} and references therein), the two cases can be discussed separately.
 
In the superconductive regime, $\mathcal{D}$ involves all the possible momenta and $\bar{\mathcal{D}} = 0$, and one has:
\begin{align}
&\frac{m}{4 \pi \hbar^2 a} = - \frac{1}{2} \,  \frac{1}{(2 \pi \hbar)^3} \int \mathrm{d} \b{k} \, \left(\frac{1}{\sqrt{\xi_{\b{k}}^2 + \Delta^2}} - \frac{1}{\varepsilon_{\b{k}}} \right) \, \label{sc_delta_ren} \\
&n= 1 - \frac{1}{(2 \pi \hbar)^3} \int_{\mathcal{D}} \mathrm{d} \b{k} \,  \frac{\xi_{\b{k}}}{E_{\b{k}}} \, , 
\label{sc_mu_ren}
\end{align}
i.e., the usual equations describing the BEC/BCS crossover, see 
\cite{randeria_rev,Zwerger12} and references therein (see as well, e.g., 
\cite{randeria} for the two-dimensional case).

As it happens for the equations (\ref{sc_finite_delta}) and (\ref{sc_finite_mu}), the parameter $\gamma$ does not appear in (\ref{sc_delta_ren}) and (\ref{sc_mu_ren}). For this reason, in the normal state, {\color{black} Eq. (\ref{gamma_cont})} can be solved independently, putting inside it the value $\tilde{\mu}$ obtained,  for  each pair of external parameters $(a, \Omega)$, from the solution of (\ref{sc_mu_ren}) {\color{black} with $\Delta = 0$}. 
Here $\bar{\mathcal{D}}= \left\{ \b{k}:  |\xi_{\b{k}}| < \Omega + \gamma \right\}=\left\{ \b{k}:  |\frac{|\b{k}|^2}{2m}- \tilde{\mu}| < \Omega + \gamma \right\}$.
The set $\bar{\mathcal{D}}$ can be rewritten as $\bar{\mathcal{D}} = \Big\{ \{\tilde{\mu} < \frac{\b{k}^2}{2m} <\tilde{\mu} + \Omega + \gamma \} \, \vee \,  \{\tilde{\mu} - (\Omega + \gamma)< \frac{\b{k}^2}{2m} <\tilde{\mu} \} \Big\}$, so that a straightforward integration of (\ref{gamma_cont}) leads to:
\beq
\gamma = - \frac{4 \sqrt{2} \, \pi}{3 \, h^3} \, U V m^{\frac{3}{2}} \Big[(\tilde{\mu} + \Omega +\gamma)^{\frac{3}{2}} - (\tilde{\mu} - \Omega- \gamma)^{\frac{3}{2}}  \Big] \, .
\eeq
This equation can be solved numerically or {\color{black} analytically} after some algebra, however the explicit solution is not very enlightening.
A considerable simplification, together with a deeper insight, can be achieved neglecting the Hartree terms from the interaction $\propto a$. {\color{black} In this way, exploiting Eq.  \eqref{relazione},
we also obtain:
\beq
\delta N = - \frac{8 \sqrt{2} \, \pi}{3 \, h^3} \, V m^{\frac{3}{2}} \Big[(\mu + \Omega)^{\frac{3}{2}} - (\mu - \Omega)^{\frac{3}{2}}  \Big] \, .
\label{umbal}
\eeq
}
 By direct inspection, it easy to check that  (\ref{umbal}) is equal to the imbalance between two free Fermi gases with chemical potential $\mu_{\pm} \equiv \mu \pm \Omega$. This result, valid also in the presence of a lattice, clarifies even more the meaning of {\color{black} $\delta N$} and $\gamma$.

\section{Further Applications}
\label{applications}
Various  applications and extensions of the Hamiltonian (\ref{H_rabi}) are interesting and experimentally feasible.
In the past Sections, we described two examples, based on imbalances in the {\color{black} initial densities of two hyperfine species in the considered mixture} or in their 
hopping terms. 
In this Section, we describe two possible further applications of the Rabi coupling, also feasible in present experiments.

\subsection{Time modulation}
An interesting set of possible applications of the Rabi coupling applied to  atomic mixtures opens {\color{black} when a time 
dependence on the time  $\tau$ is considered in Eq. \eqref{H_rabi}. 
The resulting Hamiltonian}
can be mapped by the transformation (\ref{rot}) to:
\beq
H_{\mathrm{ROT}} (\tau) = - t  \sum_{<i,j>, \alpha}  \cop{a}_{i \alpha} a_{j \alpha} 
\, -  \sum_i\left[\big( \mu - \Omega (\tau)  \big)  n_{i+}\, 
+ \,  \big(\mu  +  \Omega (\tau) \big) n_{i-}\right]
\, - \,  U \sum_{i}  n_{i+} n_{i-} \,.
\label{H_rabi_2_t}
\eeq
This Hamiltonian can be realized in the present experiments by varying in time the intensity of the lasers 
inducing the Raman transitions at the basis of the Rabi coupling.  The allowed profile of variation  is well controllable in experiments, as well as the rate 
of variation. It can be changed from scales much larger to much smaller 
than any intrinsic timescale of the experiment \cite{pagano}. In this way different situations, including quenches or adiabatic evolutions, can be probed.

Notice at the end that such a type of imbalance cannot be achieved easily without the Rabi coupling, as just acting on the numbers of atoms for each {\color{black} hyperfine species of the mixture,  since these numbers are not easily controllable in time}.

\subsection{Spatial modulations}
\label{modulated}
Another interesting extension  is the study of a space-dependence of the Rabi coupling $\Omega(i)$ in Eq. \eqref{H_rabi}.
Again, in our knowledge such a type of imbalance cannot be achieved following other techniques.
Experimentally feasible spatial dependences $\Omega(i)$  are a) gaussian: $\Omega(\b{r}, \b{r}_0) = \Omega \, e^{-\frac{|\b{r}-\b{r}_0|^2}{\sigma^2}}$, with $\sigma \approx 10-100 \, \mu \mathrm{m}$, then of the order of 10-100 lattice sites for a typical lattice; b) sinusoidal along a direction: $\Omega(\b{r}) = \Omega \, \mathrm{sin} \, \b{k}_R \cdot \b{r} $, where $\b{k}_R$ can be tuned varying the angle between the lasers inducing the Rabi coupling, and its maximum magnitude being $|\b{k}_R|^{(\mathrm{max})} = \frac{2 \pi}{\lambda_R}$ ($\lambda_R$ is the wavelength of the lasers).
In this interesting case, 
{\color{black} we obtain, via the transformation (\ref{rot}),  a spatially-modulated chemical potential:}
\beq
H_{\mathrm{ROT}} (\tau) = - t  \sum_{<i,j>, \alpha} \cop{a}_{i \alpha} a_{j \alpha} 
\,   -  \sum_{i} \left[ \big(\mu - \Omega \, \mathrm{cos} \, \vec{\phi} \cdot {\bf r}_i  \big) n_{i +}\, 
+ \, \big(\mu + \Omega  \, \mathrm{cos} \, \phi \cdot {\bf r}_i  \big) n_{i-}\right] 
\, - \,  U \sum_{i}  n_{i+} n_{i-} \, .
\label{H_rabi_s_rot}
\eeq
It would particularly interesting {\color{black} to study the effect of a generic} modulation $\vec{\phi}$ on a superfluid phase.

As addressed {\color{black} at the end of Section \ref{casoadue}}, since the Rabi coupling with spatial dependence also transfers momentum $\delta \b{p}$  to the atoms,  in the presence of an optical lattice,   its intensity is limited by the requirement
that excited Wannier functions are not populated significantly \cite{pagano}. The Rabi transition width involving different Wannier states reads $\Omega_{i, j} \propto \langle w_i(\b{r})|e^{i (\delta \b{p}) \cdot \b{r}} |w_j(\b{r})\rangle$, being $w_m(\b{r})$ the $m$-th Wannier function at each site \cite{bloch08}. Considering just the first two Wannier functions, $w_0(\b{r})$ and $w_1(\b{r})$, the probability $P_{0,1}$ of transition between them by a two-photon Raman transition giving rise to the Rabi coupling is $P_{0,1} =\Big(\frac{\Omega}{\sqrt{\Omega^2 + E_g^2}}\Big)^2$, $E_g$ being the energy difference between the two Wannier states. Imposing this quantity to be much less than 1, we obtain $\Omega << E_g$.
 For a deep optical lattice potential, where the confinement on the lattice is approximately harmonic, $E_g \sim 2 E_R \, \big(\frac{V_0}{E_R}\big)^{\frac{1}{2}}$ \cite{bloch08}, so that $\frac{\Omega}{E_R} \lesssim 2  \, \big(\frac{V_0}{E_R}\big)^{\frac{1}{2}}$. $V_0$ is the maximum intensity of the confining lattice potential and $E_R = \frac{\hbar ^ 2 k^2}{2m}$ the recoil energy, $k = \frac{2 \pi}{\lambda}$ being in turn the wave vector of the laser light creating the lattice and $m$  the mass of the atoms.
The lower bound for the ratio $\frac{V_0}{E_R}$, such that the tight-binding approximation holds, is $\frac{V_0}{E_R} \sim 5$.

The scattering amplitude $t$ for a $n$-dimensional hyper cubic lattice in the regime $\Big(\frac{V_0}{E_R}\Big) >>1$ can be estimated by 
{\color{black} as} \cite{bloch08}
\beq
t \approx \frac{4}{\sqrt{\pi}} \, E_R \, \Big(\frac{V_0}{E_R}\Big)^{\frac{3}{4}} \, \mathrm{exp}\Big[-2 \Big(\frac{V_0}{E_R}\Big)^{\frac{3}{4}} \Big] \, .
\eeq
From this formula, with $\frac{V_0}{E_R} \gtrsim 5$, 
we also obtain $\frac{t}{E_R} \lesssim 0.08$ and  $\frac{\Omega}{t} \lesssim 23.2 \,  \Big(\frac{V_0}{E_R}\Big)^{\frac{1}{2}}$. 
We find that basically the ratio $\frac{\Omega}{t}$ is only limited by the achievable intensity for $V_0$ before reaching appreciable heating regimes.
As written {\color{black} at the end of Section \ref{casoadue}}, the reachable intensities by present laser techniques can be of the same order of the Fermi energies in typical experiments, both in continuous space and on a lattice. 
{\color{black} Thus, we conclude that our method is effective in a very wide range of intensities for the Rabi coupling.}

\section{Conclusions}

\label{conclusions}

We  have shown  that a Rabi coupling applied to an atomic mixture {\color{black} of different hyperfine levels  
gives a simple method to design and control 
effective population imbalances. These effective imbalances} holds for 
dressed one-particle states obtained diagonalizing the mixing matrix related to the Rabi coupling. The method works equally for bosonic and fermionic atoms.

The presented way of controlling the chemical potential opens the possibility 
of very effectively probing the physics of interacting imbalanced mixtures. 
Indeed, as shown explicitly for {\color{black} 
a balanced and interacting two-species fermionic mixture, 
the superfluid properties in the presence of a Rabi coupling are 
the same as those for an imbalanced mixture in the absence of the Rabi term. 
{\color{black} It would be very interesting also to have spin-dependent tunnelings, in order to create 
spin-orbit-like couplings, a case that we plan to study in a future work.}

{\color{black} Notably, the proposed method   
can be also exploited to engineer spatially and/or temporally dependent 
effective population imbalances, generally not achievable in the present ultracold atoms experiments.  
We hope that the discussion presented in this paper can stimulate 
the design of new experiments in the next future.}

\acknowledgments{The authors are pleased to thank Michele Burrello, Leonardo Fallani, Francesca Maria Marchetti, {\color{black} Massimo Mannarelli}, Simone Paganelli, Guido Pagano, and Luca Salasnich for useful discussions {\color{black} and correspondence. A. C. acknowledges the support of the European Research Council (ERC) Synergy Grant UQUAM. L.D. acknowledges financial support from the BIRD2016 project "Superfluid properties of Fermi gases in optical potentials" of the University of Padova.}}

\end{document}